\documentclass[a4paper,fleqn,usenatbib]{mnras}

\usepackage[T1]{fontenc}
\usepackage{ae,aecompl}

\usepackage{graphicx}	
\usepackage{amsmath}	
\usepackage{amssymb}	
\usepackage{multirow}
\usepackage{color}
\usepackage{ulem}
\usepackage{lineno}
\title[The \textit{PoGO+} view on Crab off-pulse polarisation] {The \textit{PoGO+} view on Crab off-pulse hard X-ray polarisation}

\author[M. Chauvin et al.]{
M. Chauvin$^{1,2}$,
H.-G. Flor\'{e}n$^{3}$, 
M. Friis$^{1,2}$,
M. Jackson$^{1}$, 
T. Kamae$^{4}$, 
J. Kataoka$^{5}$,
\newauthor
T. Kawano$^{6}$
M. Kiss$^{1,2}$, 
V. Mikhalev$^{1,2}$,
T. Mizuno$^{6}$,
H. Tajima$^{7}$, 
\newauthor
H. Takahashi$^{6}$,
N. Uchida$^{6}$,
M. Pearce$^{1,2}$\thanks{E-mail: pearce@kth.se}  \\
$^{1}$KTH Royal Institute of Technology, Department of Physics, 106 91 Stockholm, Sweden\\
$^{2}$The Oskar Klein Centre for Cosmoparticle Physics, AlbaNova University Centre, 106 91 Stockholm, Sweden\\
$^{3}$Stockholm University, Department of Astronomy, 106 91 Stockholm, Sweden\\
$^{4}$University of Tokyo, Department of Physics, 113-0033 Tokyo, Japan\\
$^{5}$Research Institute for Science and Engineering, Waseda University, Tokyo 169-8555, Japan\\
$^{6}$Hiroshima University, Department of Physical Science, Hiroshima 739-8526, Japan\\
$^{7}$Institute for Space-Earth Environment Research, Nagoya University, Aichi 464-8601, Japan
}

\date{Accepted 2018 February 20. Received 2018 February 19; in original form 2018 January 7.}

\pubyear{2018}

\begin{document}
\label{firstpage}
\pagerange{\pageref{firstpage}--\pageref{lastpage}}
\maketitle

\begin{abstract}
The linear polarisation fraction and angle of the hard X-ray emission from the Crab provide unique insight into high energy radiation mechanisms, complementing the usual imaging, timing and spectroscopic approaches. Results have recently been presented by two missions operating in partially overlapping energy bands, {\it PoGO+} (18--160~keV) and {\it AstroSat} CZTI (100--380~keV). We previously reported {\it PoGO+} results on the polarisation parameters integrated across the light-curve and for the entire nebula-dominated off-pulse region. We now introduce finer phase binning, in light of the {\it AstroSat} CZTI claim that the polarisation fraction varies across the off-pulse region. Since both missions are operating in a regime where errors on the reconstructed polarisation parameters are non-Gaussian, we adopt a Bayesian approach to 
compare results from each mission. We find no statistically significant variation in off-pulse polarisation parameters, neither when considering the mission data separately nor when they are combined. This supports expectations from standard high-energy emission models.
\end{abstract}

\begin{keywords}
instrumentation: polarimeters  -- X-rays: Crab  -- methods: statistical
\end{keywords}


\section{Introduction}
\label{sec:intro}
The Crab pulsar and wind nebula are an archetypal multi-wavelength laboratory for the study of high-energy astrophysics. 
As such, the system is one of the most studied celestial objects~\citep[][]{Buhler2014}.
Additional data are required to fully understand high-energy emission processes for X-rays. 
Determining the linear polarisation of the emission provides complementary information to observation methods which result in images, energy spectra and temporal light curves~\citep[][]{Krawczynski2011}.
Linear polarisation is described using two parameters: $(i)$ the polarisation fraction (PF, \%) describing the degree of polarisation;
and, $(ii)$ the polarisation angle (PA, deg.) which describes the orientation of the electric field vector.

Polarised emission is a consequence of the synchrotron processes which are thought to dominate for the Crab~\citep[][]{Hester2008}. 
The maximum allowed PF for synchrotron emission in a uniform magnetic field is high, $\sim$75\%~\citep[][]{Lyutikov2003}.
The Crab is a complex system comprising a rotation-powered pulsar surrounded by a diffuse pulsar wind nebula which includes resolved structures in the inner nebula including a jet, toroidal structures and synchrotron shock fronts, seen as knots and 
wisps~\citep[][]{Moran2013}. 
Measured PF values are integrated over these features, and so are expected to be significantly lower than the theoretical maximum, motivating the need for sensitive and well calibrated instruments.
The dependence of the measured polarisation parameters on the rotational phase of the pulsar can be used to differentiate between emission models~\citep[][]{Cheng2000,Dyks2004,Harding2017,Petri2013}.  

More than forty~years have passed since the OSO-8 mission made the first significant observations of linear polarisation in X-ray emission from the Crab nebula. In~\citet{Weisskopf1978}, 
a relatively high PF is measured at 2.6~keV, (19.2$\pm$1.0)\% for PA=(156.4$\pm$1.4)$^{\circ}$, supporting the hypothesis that synchrotron processes dominate the emission. 
In the gamma-ray regime, inventive use of instruments on-board {\it INTEGRAL} have provided polarimetric data on the 
Crab~\citep[][]{Chauvin2013,Dean2008,Forot2008,Moran2016}.   
We note, however, that the INTEGRAL analyses are complicated by the lack of pre-launch studies of the polarimetric response of the instrument.

Recently, new results on the Crab in the hard X-ray regime (10s to 100s of keV) have been presented by two complementary polarimetry missions - the Swedish--Japanese stratospheric ballooning platform, {\it PoGO+}~\citep[][]{Chauvin2017a}, and the Indian earth-orbiting satellite, {\it AstroSat}~\citep[][]{Vadawale2017a}. 
The {\it PoGO+} mission is a development of the {\it PoGOLite Pathfinder} which detected polarisation in hard X-ray emission from the Crab in 2013~\citep[][]{Chauvin2016}.
Polarimetry measurements obtained by the CZTI instrument of {\it AstroSat} allowed the phase dependence of the polarisation parameters to be studied, whereas {\it PoGO+} estimated polarisation parameters in relatively wide phase windows.
The team analysing the {\it AstroSat} CZTI data concluded that the polarisation properties vary across the Crab off-pulse region.
This implies that the pulsar contributes significantly to the off-pulse emission. 
\citet{Vadawale2017a} point out that the viewing geometry may give rise to such effects~\citep[][]{Bai2014,Takata2007} and that a similar effect has been reported for radio pulsars~\citep[][]{Basu2011}. 
None-the-less, this is a surprising and intriguing result which warrants further study since it challenges prevailing high-energy emission models. 

In this paper, we extend our previous analysis of {\it PoGO+} Crab data by examining the phase dependence of the polarisation parameters. 
We compare our results to those obtained by {\it AstroSat} CZTI, to elucidate the off-pulse behaviour of the polarisation parameters in a lower, but partially overlapping, energy band.  

\section{New results in hard X-ray polarimetry}
\label{sec:new}
%
{\it PoGO+} (18--160~keV) is specifically designed for polarimetry, while {\it AstroSat} CZTI (100--380~keV) is a coded aperture spectrometer for general hard X-ray observations, including polarimetry. The response of both instruments was determined for both polarised and unpolarised radiation before launch~\citep[][]{Chauvin2017b,Vadawale2015}. 
Results are summarised in Table~\ref{table:comparison} and relevant characteristics of the two missions are summarised in Table~\ref{table:missions}.
In both cases, the reported values are corrected for the bias due to the positive definite nature of the 
measurement, e.g. ~\citep[][]{Maier2014}, as discussed in Section~\ref{sec:method}. 
The PF values reported are statistically compatible, although the {\it AstroSat} CZTI values are consistently higher than {\it PoGO+} which may indicate an energy dependence of PF when considered together with the OSO-8 results. 
The {\it PoGO+} PA is compatible with the pulsar spin axis, (124.0$\pm$0.1)$^{\circ}$, as determined from high-resolution {\it Chandra} X-ray images~\citep[][]{Ng2004}. The {\it AstroSat} CZTI angle is further rotated. 
When compared to OSO-8 results for the off-pulse (nebula dominated) emission, the PA does not appear to exhibit a simple evolution with energy.  
\begin{table}
\centering
\caption{Comparison of bias-corrected Crab hard X-ray polarisation parameters. Errors correspond to $1\sigma$. PA is defined relative to celestial North in the Easterly direction.}
\label{table:comparison}
\begin{tabular}{|l|c|c|c|}
\hline
{                } & {Phase range} & {PF (\%)} & {PA ($^\circ$)}  \\
\hline
\hline
{\it PoGO+}                   & All		    & 20.9$\pm$5.0 & 131.3$\pm$6.8 \\
\cline{2-4}
		  & Off-pulse	   & 17.4$^{+8.6}_{-9.3}$ & 137$\pm$15 \\		
\hline
{\it AstroSat} CZTI		 & All & 32.1$^{+5.7}_{-6.0}$ & 143.5$\pm$4.9 \\
\cline{2-4}
 		& Off-pulse & 37.7$^{+10.0}_{-11.1}$ & 141.0$\pm$7.6 \\
\hline
\end{tabular}
\end{table}

\section{Observation methodology}
\label{sec:method}
Both {\it PoGO+} and {\it AstroSat} CZTI utilise Compton scattering interactions in a segmented detector to determine the polarisation of incident X-rays. 
According to the Klein-Nishina scattering cross-section, X-rays will preferentially scatter in a direction perpendicular to the polarisation vector. 
This implies that the azimuthal scattering angle, $\phi$ (defined relative to the polarisation vector), is modulated for a given range of polar scattering angles, $\theta$.
\begin{table}
\centering
\caption{Comparison of the {\it PoGO+} and {\it AstroSat} CZTI missions (polarimetric mode).}
\label{table:missions}
\begin{tabular}{|l|c|c|c|}
\hline
{                } & {\it PoGO+} & {\it AstroSat} CZTI   \\
\hline
\hline
Platform & stratospheric balloon &  satellite \\	
Overburden & 5.8~g/cm$^2$ average & 0~g/cm$^2$ \\
\hline
Detector & plastic scintillator & CZT  \\
Pixels, geometry  & 61, hexagonal &  16384, square \\
Geometrical area   & 378~cm$^2$ & 976~cm$^2$\\
Field-of-view     & $\sim$2$^\circ$& $\sim$90$^\circ$ \\
Energy band	&  18--160~keV & 100--380~keV \\
\hline
Observation & Jul.'16 & Sep.'15--Mar.'17  \\
t$_\mathrm{source}$  & 92~ks & 800~ks\\
t$_\mathrm{bkgnd}$ & 79~ks & 180~ks \\
Signal/Bkgnd  & 0.14 & 0.05 \\
\hline
\end{tabular}
\end{table}

For {\it PoGO+}, the azimuthal scattering angle is determined from events with exactly 2 interactions at any location in the scintillator array.
The distribution of such angles is a harmonic function ("modulation curve"), where the phase defines PA, and the modulation amplitude defines PF.
In order to separate instrumental effects from source polarisation, the polarimeter is rotated around the viewing axis during observations. 
The symmetric geometry of the instrument pixels allows the scattering angle distribution to be determined independent of computer simulations.
The background is dominated by albedo atmospheric neutrons.
Since a fake polarisation signal can be generated by such an anisotropic background, 79~ks of interspersed observations  were conducted on fields 5$^{\circ}$ to the East and West of the Crab. Temporal behaviour of the background was tracked by transitioning between the fields every 15~minutes. Unbinned and background-subtracted Stokes parameters were used to determine polarisation parameters. 

Above $\sim$100~keV Compton scattering dominates in the {\it AstroSat} CZTI Cadmium-Zinc-Telluride (CZT) detector, and polarisation events are identified through coincident interactions in adjacent pixels.
The telescope structure becomes increasingly transparent across the energy range which results in a large field-of-view where, like {\it PoGO+}, observations are spatially averaged over the entire nebula.
Polarimetric data were gathered for 21 Crab observations (totalling 800~ks) after the launch on 28th September 2015. 
The low inclination (6$^\circ$) orbit provides a low background environment for measurements, with the Cosmic X-ray Background dominating.
The background response is determined through observations (180 ks) of fields with a declination close to that of the Crab, with bright X-ray sources such as the Crab and Cygnus X-1 outside the field-of-view. 
In contrast to {\it PoGO+}, the detector pixels have a square geometry which yields a non-symmetric scattering geometry. 
The CZTI instrument is not rotated during observations and uniformity corrections are derived from computer simulations. 
Resulting background-subtracted and geometry-corrected modulation curves are fit with a harmonic function to determine PA and PF.

As shown by the statistical uncertainty of the results presented in Section~\ref{sec:new}, the two missions have comparable polarimetric performance for Crab observations. The shorter observation time for {\it PoGO+} is compensated by the lower energy range (higher photon flux) and the larger Compton scattering cross-section offered by plastic scintillators. The modulation response is simpler for {\it PoGO+} due to the symmetric scattering geometry. 
Both missions report a Crab polarisation sensitivity (Minimum Detectable Polarisation, 
MDP~\citep[][]{Weisskopf2010}) before background subtraction of $\sim$10~\%. 
Unpolarised radiation has a 1\% probability of exhibiting PF$>$MDP. 
Polarisation measurements are positive definite, with the PF following a Rice distribution. This results in a bias to positive values unless the reconstructed PF$\gg$MDP.
This is not the case for either mission, so bias corrections are applied as described in the Supplementary Information of~\citet{Chauvin2017a,Vadawale2017a}.
%
\section{Phase dependent analysis of \textit{PoGO+} data}
\label{sec:results}
\begin{figure}
 \centering
    \includegraphics[width=\columnwidth]{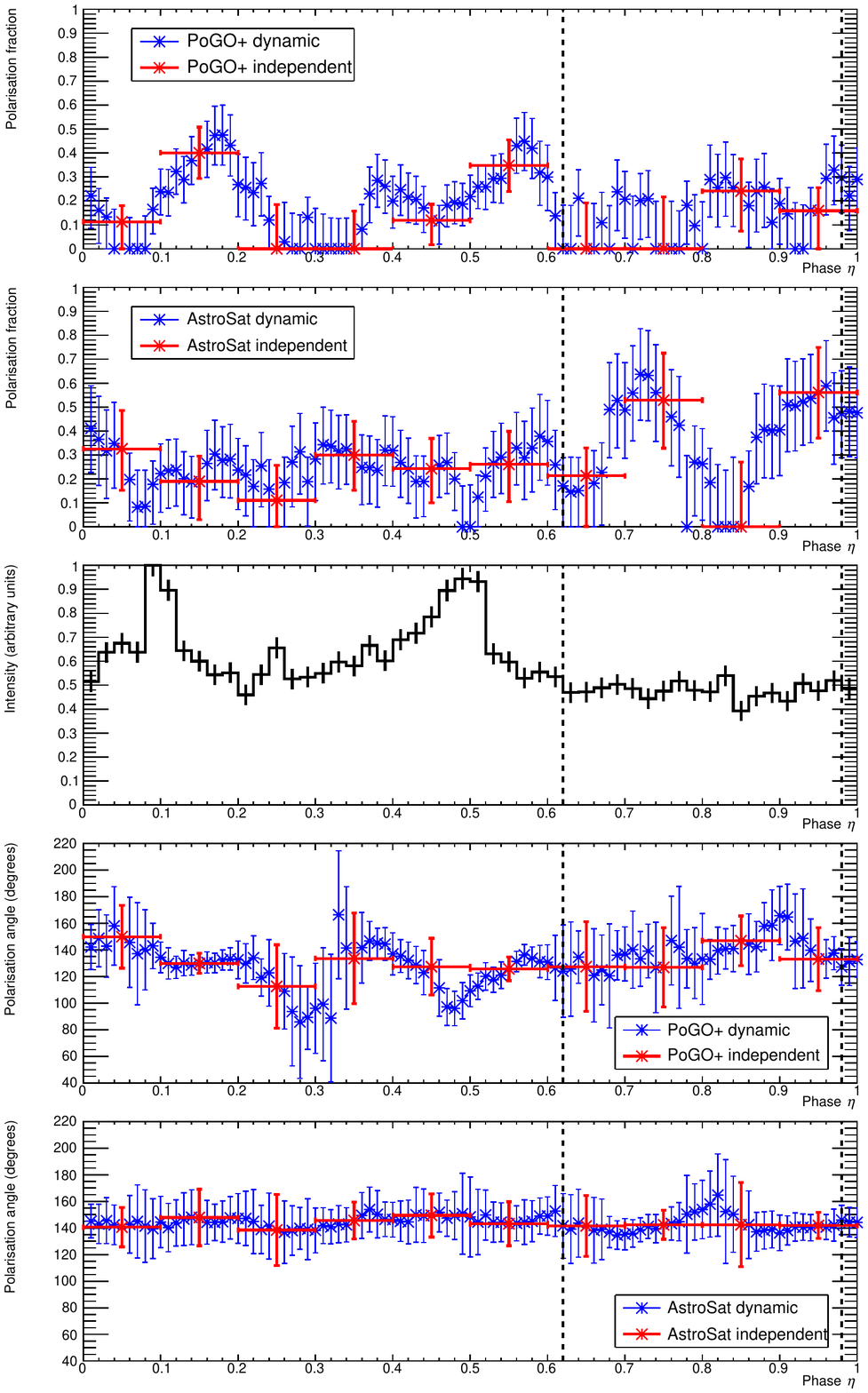}
\caption{{\bf Upper two panels:} Bias-corrected PF values, binned dynamically (blue points), and independently (red points), from {\it PoGO+} and {\it AstroSat} CZTI data. 
{\bf Middle panel:} The background-subtracted {\it PoGO+} Crab light-curve (shown for reference) with the off-pulse region bounded by dashed lines. 
{\bf Lower two panels:}  Bias-corrected PF values, binned dynamically (blue points), and independently (red points), from {\it PoGO+} and {\it AstroSat} CZTI data.
The binning convention follows that presented in~\citet{Vadawale2017a}. The {\it AstroSat} CZTI error estimates are revised, as described in Section~\ref{sec:results}.
}
  \label{fig:combined}
\end{figure}

In~\citet{Chauvin2017a} we determined polarisation parameters integrated over the off-pulse region. We now extend this work by following the "dynamic binning" approach detailed in~\citet{Vadawale2017a} in order to directly compare with {\it AstroSat} CZTI. 
Polarisation parameters are determined across the full phase range, $0\leq \eta \leq1$, in bins of width 0.1, spaced with a phase interval of 0.01. 
Results are shown in Fig.~\ref{fig:combined}. 
The error bars shown for the {\it AstroSat} CZTI data are different to those presented in~\citet{Vadawale2017a} (Figure 7, Supplementary Information) since confidence levels where inadvertently shown instead of credibility intervals~\citep[][]{Vadawale2017b}.
For the majority of the phase bins the difference is negligible but low significance bins were reported with over-estimated errors, e.g. for $\eta=0.83$ the credibility interval should be [0\%,23.6\%] not [0\%,29.8\%].
The interpretation of polarisation parameter trends from this binning approach is complicated by the presence of correlations - these figures therefore also show parameters derived for 10 independent phase bins, selected to follow the {\it AstroSat} CZTI convention. 
For the PF data, {\it PoGO+} indicates more variation in the pulsar peaks (where pulsar and nebula emission are mixed), while {\it AstroSat} CZTI data varies more in the off-pulse region. 
This off-pulse variation is stated as a main result of the  {\it AstroSat} CZTI analysis.
In the PA data, no significant variation is seen in either sets of data.
In the remainder of this paper, we present a statistical analysis on the three independent phase-bins in the off-pulse region where variation in PF is reported by {\it AstroSat} CZTI.
%

\section{Statistical comparison to \textit{AstroSat} CZTI results}
\label{sec:statistics}
\begin{figure}
 \centering
    \includegraphics[width=\columnwidth]{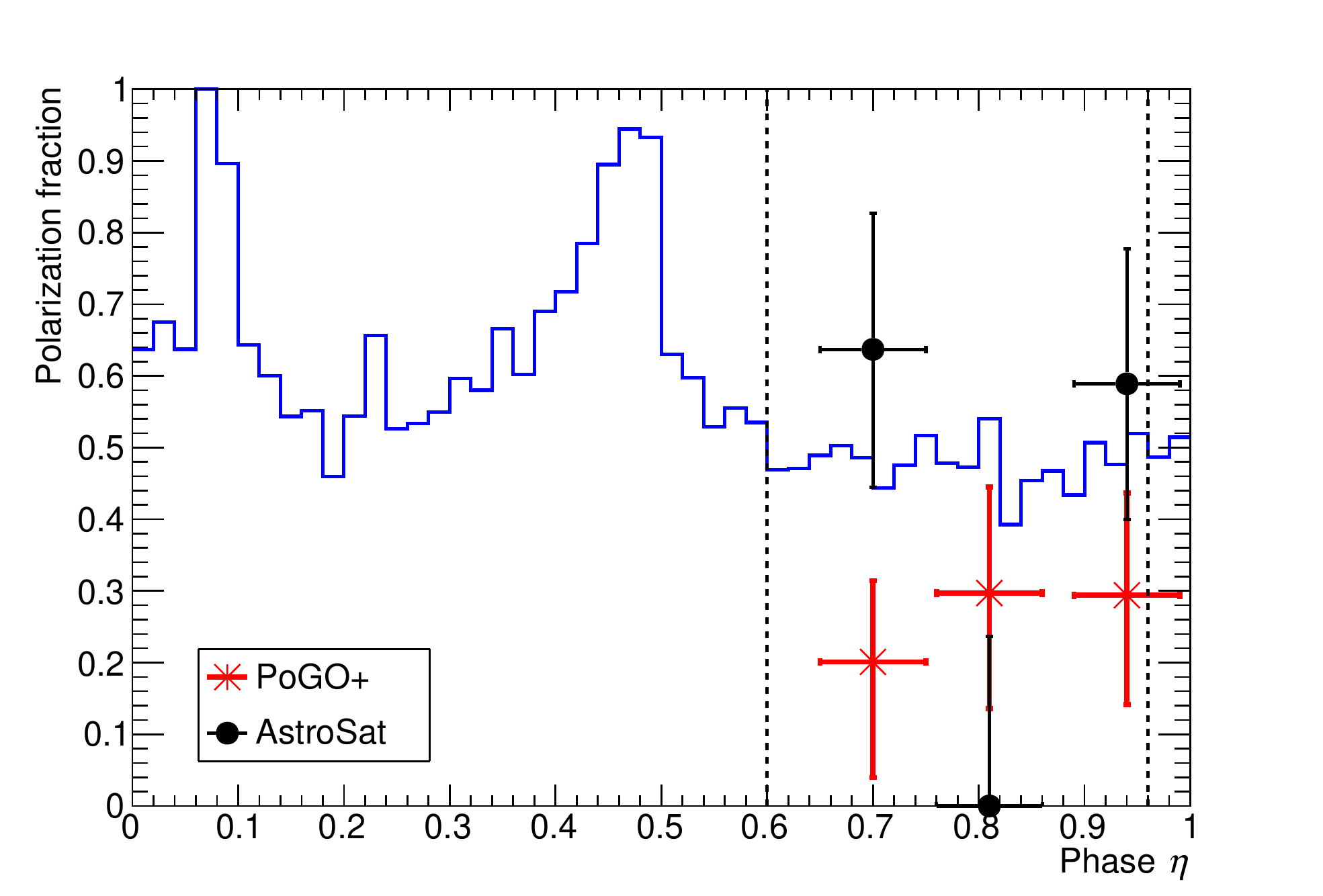}
\caption{PF in the off-pulse region displayed in three independent phase bins, as selected by~\citet{Vadawale2017a}. {\it PoGO+} ({\it AstroSat} CZTI) data are shown as red (black) data-points. The light-curve reconstructed by {\it PoGO+} is shown in blue. Compared to Fig.~\ref{fig:combined}, the light-curve has been shifted by 0.02 in phase in order to accommodate the width of the data-points. The {\it AstroSat} CZTI error estimates are revised, as described in Section~\ref{sec:results}.} 
  \label{fig:superimposed}
\end{figure}

We have studied the statistical significance of variations in polarisation parameters across the Crab off-pulse phase region, using the independently binned PF results shown in Fig.~\ref{fig:superimposed}. 
A standard $\chi^2$ approach is not suitable since PF$\sim$MDP, which yields non-Gaussian errors. We instead use a Bayesian approach in our analysis.  
%
\subsection{Bayesian model comparison}
The Bayesian methodology provides an automatic way of applying Occam's razor to model selection~\citep[][]{Trotta2007}. 
An important aspect is that under-utilised data-set space, which favours simple models, is automatically balanced against differences between predicted and observed data, which favours complex models. 

The posterior distribution over the models $\mathcal{M}_i$ for all data $\mathcal{D}$ is given by
\begin{equation}
P(\mathcal{M}_i|\mathcal{D})\propto P(\mathcal{D}|\mathcal{M}_i)P(\mathcal{M}_i),
\end{equation}
where, $P(\mathcal{M}_i)$ is a prior over the models and $P(\mathcal{D}|\mathcal{M}_i)$ is the evidence (the marginal likelihood),
\begin{equation}
P(\mathcal{D}|\mathcal{M}_i)=\int P(\mathcal{D}|\mathbf{w},\mathcal{M}_i)P(\mathbf{w}|\mathcal{M}_i)\mathrm{d}\mathbf{w}.
\end{equation}
Here, $\mathbf{w}$ is a set of parameters, $P(\mathbf{w}|\mathcal{M}_i)$ is the prior over parameters given a model and $P(\mathcal{D}|\mathbf{w},\mathcal{M}_i)$ is the likelihood (explicit form given in Appendix~\ref{appendix:likelihood}). When comparing two models, the Bayes factor is a ratio of the evidences,
\begin{equation}
B_{01}=\frac{P(\mathcal{D}|\mathcal{M}_0)}{P(\mathcal{D}|\mathcal{M}_1)}
\end{equation}
and can be interpreted as an odds ratio between models when their priors are equal i.e. $P(\mathcal{M}_0)=P(\mathcal{M}_1)=0.5$. In that case the probability for $\mathcal{M}_1$ given that there are only two possible models is
\begin{equation}
P(\mathcal{M}_1|D)=\frac{1}{1+B_{01}}.
\end{equation}
\subsection{Models and priors}
We follow a parametric approach to quantify statistically the claim by \cite{Vadawale2017a} that PF varies in the off-pulse region.
The function $f(\eta;\mathbf{w})$ (see Appendix~\ref{appendix:likelihood}) should contain few parameters since there are only three data-points in Figure~\ref{fig:superimposed}.
A Bayesian model comparison where there are more parameters than data-points is possible, however such models will be penalized since there is only unit probability density to distribute among all possible datasets, $\mathcal{D}$.

\begin{figure*}
 \centering
    \includegraphics[width=12cm]{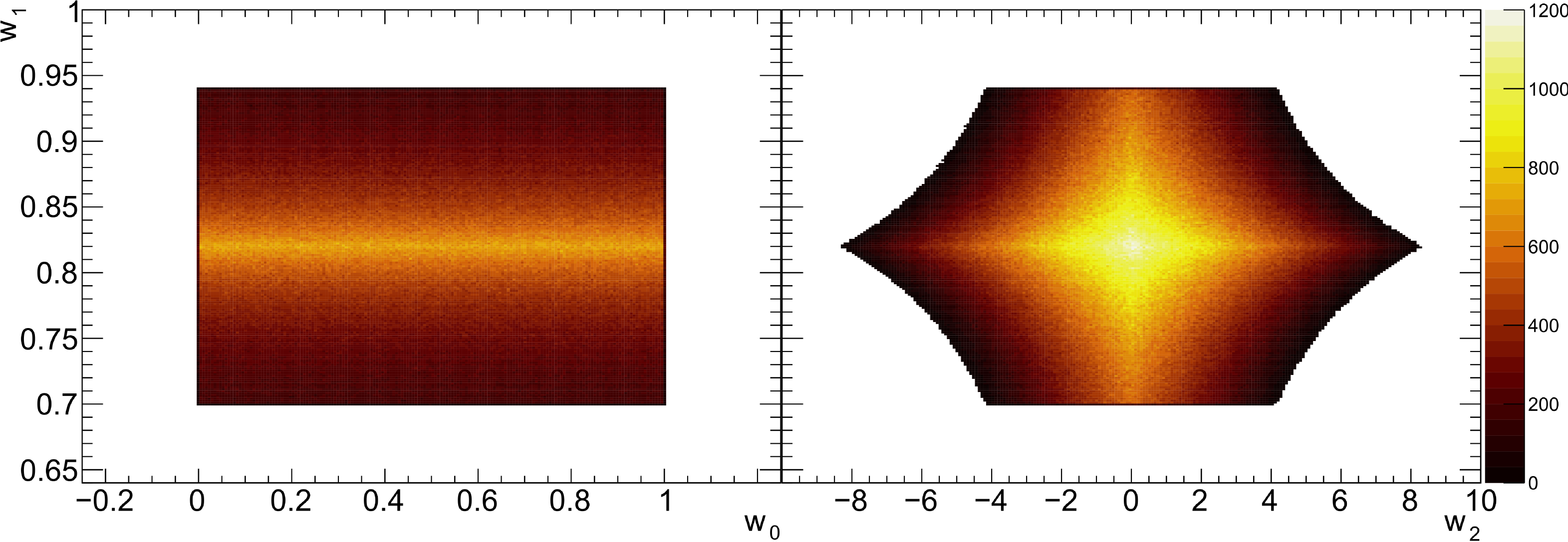}
\caption{Prior distributions for the weights $\mathbf{w}$. The priors are constructed by drawing uniform random numbers from a large range and then excluding $\mathbf{w}$ where $f_i(\eta;\mathbf{w})<0$ or $f_i(\eta;\mathbf{w})>1$ for $0.70\leq\eta\leq0.94$. The number of samples is shown on z-axis.} 
  \label{fig:priorcombined}
\end{figure*}

We consider the models in Table~\ref{tab:models} where $\mathcal{M}_0$ corresponds to no change in PF across the off-pulse region.
The other models, $\mathcal{M}_{1-4}$, are V-shaped functions with different parameter constraints.  
The parameters, $w_0$, $w_1$ and $w_2$, correspond to an offset from zero, the phase of the extreme point and the gradient, respectively. 
Model $\mathcal{M}_4$ is, arguably, the most physical since it requires the change to occur at the same phase and in the same direction for the partially overlapping energy bands of {\it PoGO+} and {\it AstroSat} CZTI. However, a general approach is followed and different weights are allowed under the same model unless specified in the "shared traits" column.
A simple first-order polynominal is not considered since it has significantly lower evidence for both missions. 
\begin{table}
\centering
\caption{The models and functions used in the likelihood $P(\mathcal{D}|\mathbf{w},\mathcal{M}_i)$ as specified by Eq.~\ref{eq:likelihood}, where $\eta$ is the pulsar phase. The "shared traits" column indicates which parameters are forced to be the same for {\it PoGO+} and {\it AstroSat} CZTI data.}
\label{tab:models}
\begin{tabular}{|l|c|c|}
\hline
{Model} & {Function $f_i(\eta)$ } & {Shared traits}  \\
\hline \hline 
$\mathcal{M}_0$			& $w_0$ & None \\
$\mathcal{M}_1$ 			& $w_2|\eta-w_1|+w_0$ & None \\
$\mathcal{M}_2$ 			& $w_2|\eta-w_1|+w_0$ & $w_1$ \\
$\mathcal{M}_3$			& $w_2|\eta-w_1|+w_0$ & sign($w_2$) \\
$\mathcal{M}_4$ 			& $w_2|\eta-w_1|+w_0$ & $w_1$ \& sign($w_2$) \\
\hline
\end{tabular}
\end{table}
The parameters, $w_0$, $w_1$ and $w_2$, are chosen from uniform distributions but not all combinations are possible as they would yield unphysical results, e.g. PF$<$0 or PF$>$1. 
Instead of determining the inter-dependence of parameters, the parameters are sampled randomly, discarding combinations that are non-physical. After sufficiently many iterations, the entire valid parameter space is sampled. This results in the priors shown in Figure~\ref{fig:priorcombined}.

\section{Results and discussion}
The Bayes factors and evidence values (Appendix~\ref{appendix:evidence}) are shown in Table~\ref{tab:separate} for each mission. The Bayes factors are close to unity, so the data provides little information about the phase evolution of PF. As expected, {\it PoGO+} tends to favour the constant model $\mathcal{M}_0$, while {\it AstroSat} CZTI favours the V-shape $\mathcal{M}_1$.
The Bayes factors and evidence values when combining data from {\it PoGO+} and {\it AstroSat} CZTI
are shown in Table~\ref{tab:combined}. The Bayes factors are very close to unity and there is no clear separation between the 
models - independent of model choice. Consequently, we do not support the claim that there is a variation in polarisation 
properties within the off-pulse region.
While a re-flight of the {\it PoGO+} mission is not foreseen, additional data from {\it AstroSat} CZTI may help to clarify the situation.  

\begin{table}
\centering
\caption{Mission-wise model comparison.}
\label{tab:separate}
\begin{tabular}{|l|c|c|c|c|}
\hline
	& {$P(\mathcal{D}|\mathcal{M}_0)$} & {$P(\mathcal{D}|\mathcal{M}_1)$} & {$B_{01}$}  & $P(\mathcal{M}_1|D)$ \\
\hline \hline 
{\it PoGO+}			& 3.88 & 2.39 & 1.62 & 0.38 \\
{\it AstroSat} CZTI		& 0.34 & 0.70 & 0.48 & 0.68 \\
\hline
\end{tabular}
\end{table}
\begin{table}
\centering
\caption{Model comparison for combined data from {\it PoGO+} and {\it AstroSat} CZTI.}
\label{tab:combined}
\begin{tabular}{|l|c|c|c|}
\hline
{Model} & {$P(\mathcal{D}|\mathcal{M}_i)$} & {$B_{0i}$}  & $P(\mathcal{M}_i|D)$\\
\hline \hline 
$\mathcal{M}_0$			& 1.31 & - & - \\
$\mathcal{M}_1$ 			& 1.69 & 0.78 & 0.56 \\
$\mathcal{M}_2$ 			& 1.83 & 0.72 & 0.58 \\
$\mathcal{M}_3$			& 1.46 & 0.90 & 0.53 \\
$\mathcal{M}_4$ 			& 1.50 & 0.88 & 0.53 \\
\hline
\end{tabular}
\end{table}
%
\section*{Acknowledgements}
We are grateful to S.~Vadawale and the {\it AstroSat} CZTI Collaboration for generously making their data available to us and for valuable discussions.
This research was supported by The Swedish National Space Board, The Knut and Alice Wallenberg Foundation, The Swedish Research Council, The Japan Society for Promotion of Science and ISAS/JAXA. 









\appendix

\section{Likelihood}
\label{appendix:likelihood}
PF is a positive definite quantity -- an unpolarised source will always have a non-zero reconstructed polarisation $p_r$. The likelihood for observing $N$ data-points with PF values $p_{r_j}$, after marginalising over the polarisation angle (which is assumed constant) is a product of Rice distributions,
\begin{equation}
\begin{split}
P(\mathcal{D}|\mathbf{w},\mathcal{M}_i)=\prod_{j=1}^{N}\frac{p_{r_j}}{\sigma^2_j}\exp{\bigg(-\frac{p_{r_j}^2+f_i(\eta;\mathbf{w})^2}{2\sigma^2_j}\bigg)} \\
\times I_0\bigg(\frac{p_{r_j}\times f_i(\eta;\mathbf{w})}{\sigma^2_j}\bigg)
\end{split}
\label{eq:likelihood}
\end{equation}
where $I_0$ is a modified Bessel function of the zeroth order, $\eta$ is the pulsar phase, $p_0=f(\eta;\mathbf{w})$ is the true PF value parameterised with weights $\mathbf{w}$, and $\sigma$ is the effective uncertainty on PF. Note that $\sigma$ does not correspond to a Gaussian sigma.
Since the signal-to-background ratio $\mathcal{R}=S/B$ is low for both {\it PoGO+} and {\it AstroSat} measurements and the calibration parameters have insignificant uncertainties, i.e. $\mu_r=\mu_0$, it is possible to write
\begin{equation}
\sigma=\frac{2}{\mu_r}\sqrt{\frac{1}{S}\bigg(\frac{S+B}{2S}-\frac{\mu_0^2p_0^2}{4}\bigg)}=\frac{\sqrt{2(S+B)}}{\mu_r S},
\end{equation}
removing the dependency on $p_0$. This makes $\sigma$ a fixed parameter for every data-point since it does not depend on the polarisation properties but only on the signal and background counts which have insignificant uncertainties,
thus $\mathcal{D}=\{p_{r_1},...,p_{r_N};\sigma_1,...\sigma_N\}$.
%
\section{Computing the evidence}
\label{appendix:evidence}

We compute the evidences $P(\mathcal{D}|\mathcal{M}_i)$ using the Monte Carlo integration
\begin{equation}
P(\mathcal{D}|\mathcal{M}_i)=\lim_{S\to\infty} \frac{1}{S}\sum_{j=1}^S P(\mathcal{D}|\mathbf{w}_j,\mathcal{M}_i),
\end{equation}
where $\mathbf{w}_j$ is randomly sampled from the prior $P(\mathbf{w}|\mathcal{M}_i)$ and $S$ is the number of samples. We use $S=10^7$ which is sufficient for a good approximation of the evidence.


\bsp	
\label{lastpage}
\end{document}